\documentclass[aps,prd,showpacs,amsmath,preprint]{revtex4}
\usepackage[dvips]{color,graphicx}
\usepackage{amsfonts}
\usepackage{amssymb}
\usepackage{latexsym}

\newcommand{\dalm}{\kern1pt\vbox{\hrule height 0.9pt\hbox{\vrule width
0.9pt\hskip 2.5pt\vbox{\vskip 5.5pt}\hskip 3pt\vrule width 0.3pt}\hrule height
0.3pt}\kern1pt}
\newcommand{\ma}[1]{\mbox{$\mathcal{#1}$}}

\newcommand{\lw}[1]{\smash{\lower2.ex\hbox{#1}}}

\begin{document}


\title{Origin of matter out of pure curvature}{
\author{
Naresh Dadhich$^{(1)}$\footnote{Electronic address:nkd@iucaa.ernet.in}
and Hideki Maeda$^{(2),(3),(4),(5)}$\footnote{Electronic address:hideki@gravity.phys.waseda.ac.jp}
}

\affiliation{
$^{(1)}$Inter-University Centre for Astronomy \& Astrophysics, Post Bag 4, Pune~411~007, India\\
$^{(2)}$Centro de Estudios Cient\'{\i}ficos (CECS), Casilla 1469 Valdivia, Chile\\
$^{(3)}$Graduate School of Science and Engineering, Waseda University, Tokyo 169-8555, Japan\\
$^{(4)}$Department of Physics, Rikkyo University, Tokyo 171-8501, Japan\\
$^{(5)}$Department of Physics, International Christian University, 3-10-2 Osawa, Mitaka-shi, Tokyo 181-8585, Japan
}
\date{\today}


\begin{abstract}

We propose a mechanism for origin of matter in the universe in the
framework of Einstein-Gauss-Bonnet gravity in higher dimensions.
The recently discovered new static black hole solution by the
authors~\cite{md2006} with the Kaluza-Klein split up of spacetime as a
product of the usual ${\ma M}^4$ with a space of negative constant
curvature is indeed a pure gravitational creation of a black hole which is
also endowed with a Maxwell-like {\it gravitational charge} in
four-dimensional vacuum spacetime. Further it could be envisioned as being
formed from anti-de Sitter spacetime by collapse of radially inflowing
charged null dust. It thus establishes the remarkable reciprocity between
matter and gravity - as matter produces gravity (curvature), gravity too
produces matter.
\end{abstract}
 

\maketitle

\clearpage

Einstein's theory of gravitation is characterized by its distinctive feature of gravity being described by curvature of spacetime. 
Matter produces  gravity; i.e. curvature in spacetime. On the other way round, reciprocity of action between matter 
and curvature (gravity) will demand that the latter should also produce the former.  
However except for Kaluza-Klein theory there is no direct demonstration of this feature. 
The nearest we have come in the string theoretic formulation is what is known as AdS/CFT correspondence~\cite{mal}.  
In here, at the boundary of anti-de~Sitter (AdS) spacetime there resides conformal field theory (quantum chromodynamics) of matter.  AdS is pure curvature while CFT is matter field and there exists a duality correspondence between them. 
This is indeed the first example of spacetime curvature manifesting itself as matter in a dual spacetime.

 Very recently, we have obtained an interesting $n(\ge 6)$-dimensional vacuum black hole solution of Einstein-Gauss-Bonnet gravity in Kaluza-Klein spacetime with extra dimensional space having negative constant curvature~\cite{md2006}. 
The four-dimensional spacetime in the solution is asymptotically Reissner-Nortstr\"om-AdS (RN-AdS) spacetime in spite of absence of Maxwell field.
Indeed, this offers a direct and purely classical example of curvature manifesting as matter, i.e., ``matter without matter''.
The key ingredient of this remarkable result is the Kaluza-Klein decomposition which also brings Gauss-Bonnet contribution down to four dimensions. 

Higher dimensions greater than usual four is the natural playground for superstring/M-theory~\cite{superstring,Lukas}. 
However classically since gravity is to be described by curvature of spacetime, that means its Lagrangian should be an invariant function of Riemann curvature and its contractions. 
The equation following from its variation should be second order quasi-linear (highest order of derivative to be linear) differential equation so that initial value problem is well formulated and evolution is uniquely defined. 
This uniquely leads to the Lovelock polynomial~\cite{lovelock}, ${\cal L}_0 = -2\Lambda,~{\cal L}_1 = R, {\cal L}_2 =L_{GB} \equiv  R^2 - 4 R_{\mu\nu}R^{\mu\nu} + R_{\mu\nu\rho\sigma}R^{\mu\nu\rho\sigma},~...$ as the general gravitational action. 
The first two are the familiar cosmological constant and Einstein-Hilbert action while the quadratic is Gauss-Bonnet term. 
For $n \le 4$, the first two suffice as higher order terms make no contribution to the equation of motion. 
This means the order of Lagrangian to be included for description of classical gravity depends upon spacetime dimension. 
It is though true that our experience is restricted to four dimension and all empirical and experimental support for gravitational dynamics emanates only from that. 
The consistency of principle and concept would however demand that for $n \ge 5$, higher order terms in the Lovelock polynomial are also equally valid and should be included for full description of gravitational field. 

It turns out that the low-energy limit of heterotic superstring theory as the higher curvature correction to general relativity~\cite{Gross} indeed gives rise to quadratic Gauss-Bonnet term $L_{GB}$. 
On the other hand one of the present authors has argued quite forcefully that there are also purely classical and strong physical motivations for higher dimension~\cite{dad}. 
For example, keeping gravity confined to $4$ dimension is equivalent to confining spacetime curvature entirely to $4$-spacetime. 
The question is, how do we ensure that? 
That is no curvature information is transmitted to extra dimension, and for that to happen $4$-spacetime should be isometrically embeddable in five-dimensional flat spacetime. 
This is however not the case in general and there is a well known theorem in differential geometry which states that an arbitrarily curved $4$-space requires six ($=n(n-1)/2$ with $n=4$) extra dimensions for its flat space embedding~\cite{spivak}. 
Gravity could thus penetrate down to ten dimension! 
Also gravity is self interactive and self interaction can be evaluated iteratively. 
The first iteration is however already included in Einstein's equation as it contains square of first derivative of the metric. 
The question is, how do we stop at the first iteration? 
We should go to second, third and so on. 
It turns out that the terms in the Lovelock polynomial represent the iterations with linear (Einstein-Hilbert) containing the first while the quadratic (Guass-Bonnet) the second and so on. 
But Gauss-Bonnet term makes no contribution in the equation for $n\le 4$ and hence we have to go to higher dimension to physically realize the second iteration of self interaction. 
As $2$ and $3$ dimensions are not big enough to accommodate free gravity, similarly $4$ dimension is not big enough to fully accommodate self interaction of gravity. 
Thus even classical dynamics of gravity requires higher dimension.   

We write action for $n$-dimensional spacetime, 
\begin{equation} 
\label{action}
S=\int d^nx\sqrt{-g}\biggl[\frac{1}{2\kappa_n^2}(R-2\Lambda+\alpha{L}_{GB}) \biggr]+S_{\rm matter},
\end{equation}
where $R$ and $\Lambda$ are $n$-dimensional Ricci scalar and the cosmological constant respectively. 
Further $\kappa_n\equiv\sqrt{8\pi G_n}$, where $G_n$ is $n$-dimensional gravitational constant and $\alpha$ is Gauss-Bonnet coupling constant.
In string theoretic formulation, $\alpha$ is identified with the inverse string tension and is positive definite~\cite{Gross}. 
We shall therefore take $\alpha \ge 0$.

The gravitational equation following from the action (\ref{action}) is given by 
\begin{equation}
{G}^\mu_{~~\nu} +\alpha {H}^\mu_{~~\nu} +\Lambda \delta^\mu_{~~\nu}=\kappa_n^2 {T}^\mu_{~~\nu}, \label{beq}
\end{equation}
where ${G}_{\mu\nu}$ is the Einstein tensor and  
\begin{eqnarray}
{H}_{\mu\nu}&\equiv&2\Bigl[RR_{\mu\nu}-2R_{\mu\alpha}R^\alpha_{~\nu}-2R^{\alpha\beta}R_{\mu\alpha\nu\beta} \nonumber \\
&&~~~~+R_{\mu}^{~\alpha\beta\gamma}R_{\nu\alpha\beta\gamma}\Bigr]
-{1\over 2}g_{\mu\nu}{L}_{GB}.
\end{eqnarray}
$T_{\mu\nu}$ is energy-momentum tensor of matter field derived from $S_{\rm matter}$ in the action (\ref{action}). 
It is noted that Gauss-Bonnet term makes no contribution in the field equation, i.e. $H_{\mu\nu} \equiv 0$, for $n \le 4$.

We shall now demonstrate creation of black hole from constant curvature of extra dimensional space by considering a six-dimensional Kaluza-Klein vacuum spacetime which is locally homeomorphic to ${\ma M}^4 \times {\ma H}^2$ with the metric $g_{\mu\nu}=\mbox{diag}(g_{AB},r_0^2\gamma_{ab})$, $A,B = 0,..,3;~a,b = 4,5$. Here $g_{AB}$ is an arbitrary Lorentz metric on ${\ma M}^4$, $r_0$ is a constant and $\gamma_{ab}$ is the unit metric on the two-dimensional space of negative constant curvature ${\ma H}^2$.
Then vacuum equation ($T_{\mu\nu}=0$) gets decomposed as follows ~\cite{md2006}: $C_1 G_{AB} +C_2 g_{AB} = 0$ and $F g_{ab} = 0$ where $C_1 \equiv 1 - 4\alpha/r_0^2$, $C_2 \equiv \Lambda +1/r_0^2$ and $F \equiv \Lambda - (R + \alpha L_{GB})/2$. Note that all curvature quantities refer to ${\ma M}^4$ and for arbitrary $C_1$ and $C_2$, the first equation is the usual Einstein equation with $\Lambda$. 
For spherically symmetric metric in the areal coordinates, it is clearly the Schwarzschild-(A)dS solution. The scalar constraint $F = 0$ will then require square of Riemann curvature, Kretshmann scalar to be constant. That could only happen if the Schwarzschild mass vanishes leaving spacetime to be (A)dS. Thus scalar constraint following from extra-dimensional equation does not let $4$-spacetime to harbour any mass point. This is an important result. 

However there is an exceptional case of $C_1 = C_2 = 0$, which means 
$r_0^2 = 4\alpha = -1/\Lambda $, and then $4$-metric remains completely free and undetermined. 
It is determined entirely by the scalar constraint $F = 0$, which is given by
\begin{eqnarray}
R+\alpha L_{GB}+\frac{1}{2\alpha}=0. \label{KKbasic}
\end{eqnarray}
For spherically symmetric spacetime in the Schwarzshild gauge, $g_{tt}g_{rr} = -1$, this admits the general solution as 
\begin{equation}
f(r)\equiv -g_{tt}=1+\frac{r^2}{4\alpha}\biggl[1\mp\biggl\{\frac{2}{3}+\frac{16\alpha M}{r^3}-\frac{16\alpha q}{r^4}\biggl\}^{1/2}\biggl], \label{special}
\end{equation}
where $M$ and $q$ are constants of integration. 
This is the new black-hole solution~\cite{md2006} for the upper negative sign and we take $q < 0$ so that the expression under the radical sign remains always positive. Asymptotically it approximates to Reissner-Nordstr\"om charged black hole in AdS spacetime even though there was no Maxwell field introduced in ${\ma M}^4$. $q$ is new ``gravitational charge'' like Weyl charge in the brane world gravity~\cite{dmpr00}. 
In four dimension, Maxwell field is characterized by $T = 0$. The scalar constraint also implies vanishing of trace and that is why gravitational charge, $q$ resembles Maxwell charge~\cite{next}. This Maxwell-like field is thus pure gravitational creation. Also note that the 
solution has no general relativistic limit, $\alpha\to0$, which is indicative of the fact that its source is entirely the quadratic Gauss-Bonnet term. 

It is also remarkable that this solution could be generalized to include radially flowing Vaidya radiation, as in the Schwarzshild case, simply by making the parameters $M = M(v),~q = q(v)$ arbitrary functions of advanced time $v$. 
This will asymptotically resemble charged null dust~\cite{next}. Why null dust because it also has $T=0$. That is only trace free matter could be created by this prescription. Both the solutions have metric regular everywhere, though curvatures diverge but weakly as $1/r^2$ or $1/v^2$. 
If we integrate the Kretshmann scalar over the volume, it will vanish as $r,v \to 0$. 
Singularity is thus weakened which is also the case for five-dimensional Gauss-Bonnet black hole~\cite{GBBH,dad}. 
It is interesting that by Kaluza-Klein construction we have been able to bring this desirable feature of Einstein-Gauss-Bonnet gravity down to four dimension. Further, it may be noted that the asymptotic limit of the solutions is not flat but AdS. This is however clear from the equation (\ref{KKbasic}) that it cannot admit any solution with asymptotic flat limit. In Gauss-Bonnet gravity, there are two branches of the solution, one admitting $\alpha\to0$ limit with flat asymptotic and the other with no $\alpha\to0$ limit with AdS asymptotic ~\cite{GBBH,dad}. Note that in our case both branches of the solution (\ref{special}) have AdS asymptotic. Thus, what flat space is to Einstein gravity, AdS is to Gauss-Bonnet gravity in Kaluza-Klein spacetime. 

Our main aim in this essay is to demonstrate creation of matter out of curvature - a pure gravitational creation as an example of {\it matter without matter}. The scalar constraint as well as the prescription ($C_1 = 0, C_2 = 0$) of extra dimensional curvature and $\Lambda$ in terms of Gauss-Bonnet parameter $\alpha$ entirely prohibit presence of any matter $T_{AB}$ on $4$-spacetime. A ``charged'' black hole and null dust are therefore pure gravitational creatures born out of coupling of curvature of extra dimensional space with $\Lambda$ and  $\alpha$. We thus establish the basic physical principle of reciprocity between gravity and matter. Using these solutions, we can construct the four-dimensional spacetime representing a black-hole formation from AdS spacetime without any matter field as shown in Fig. 1. It could thus be envisioned that static black hole is created from AdS spacetime by implosion of radially infalling Vaidya-like charged null dust. 

However one may wonder what created the parameters $M$ and $q$? Clearly their source whatever may it be cannot reside on ${\ma M^4}$. They arose from integration of the scalar constraint which involved scalar curvature and the quadratic $L_{GB}$. In absence of the latter, it would give RN-AdS and then there won't be any prohibition on existence of stress tensor which could source mass and charge parameters. In our case, we have the only one equation (\ref{KKbasic}) in which $L_{GB}$, as pointed out earlier, acts as source and there can exist no other matter stress distribution. Thus the source for these parameters can't be anything other than  $L_{GB}$, the quadratic in curvatures, which as argued earlier represents second iteration of self interaction. Thus it is the second order linking of gravity with itself which is the source of this interesting black hole spacetime. This is how it is truly a pure curvature creation, a remarkable example of matter without matter.

\begin{figure}[tbp]
\includegraphics[width=.40\linewidth]{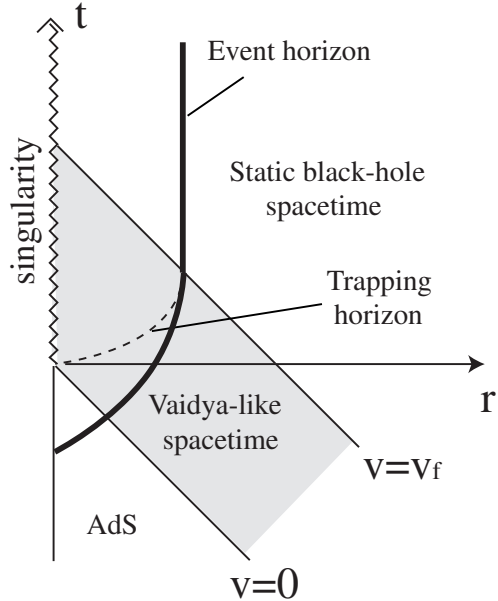}
\caption{
A schematic diagram of the black-hole formation from the AdS spacetime without any matter field.
Here the AdS spacetime for $v<0$ is joined to the static black-hole spacetime (\ref{special}) for $v>v_{\rm f}$ by way of the Vaidya-like spacetime.
We set $M(v)=m_0v$ and $q(v)=-\alpha$, where $m_0$ is a positive constant.
A singularity is formed at $v=r=0$ and develops.
}
\label{Fig1}
\end{figure}

The first realization of the principle of reciprocity was experimentally observed in the electron-positron pair creation from radiation which did require magnetic field as the facilitating agent. Similarly here we need extra dimensions and Kaluza-Klein split up of 
spacetime for creating ``matter'' from curvature. It is this split up together with Gauss-Bonnet term and $\Lambda$ which first allowed a prescription of parameters such that presence of matter was prohibited and then it provided a scalar equation which solved to give the black hole. This means that reciprocity seems to require specific prescription. So is also the case for AdS/CFT correspondence where gravity lives in bulk spacetime while matter resides on its boundary. Thus decomposition of spacetime into two is the facilitating mechanism for this process.  

In the original Kaluza-Klein theory, origin of Maxwell field is the extra-dimensional component of $5$-dimensional metric for which $5$-dimensional vacuum Einstein equation is decomposed into $4$-dimensional Einstein-Maxwell equation~\cite{ow1997}.
Here we have given a completely different and novel origin for Maxwell-like field as well as a null dust fluid in the framework of Einstein-Gauss-Bonnet gravity. 
This is however a partial success as it can only create trace free matter and the question still remains for origin of matter with non-zero trace. This is an important question and its solution will perhaps have profound impact on our understanding of spacetime and matter, and the universe.

\acknowledgments
We thank Hideki Ishihara, Panagiota Kanti, Umpei Miyamoto, Masato Nozawa and Rong-Gen Cai for comments and Albert Einstein Institute, Golm for hospitality which facilitated this work.  
HM was supported by the Grant-in-Aid for Scientific Research Fund of the Ministry of Education, Culture, Sports, Science and Technology, Japan (Young Scientists (B) 18740162).


\end{document}